\documentclass{article}
\usepackage{geometry}
\usepackage[hidelinks]{hyperref}
\usepackage{xcolor}

\usepackage{chemformula}
\usepackage{siunitx}
\usepackage{booktabs}
\usepackage{makecell}
\usepackage{newtxtext, newtxmath}
\usepackage[scaled=0.9]{helvet}
\usepackage{microtype}
\usepackage[bf, sf]{titlesec}
\usepackage[labelfont=bf]{caption}
\usepackage[
	backend=biber,
	style=numeric-comp,  % Inline square brackets and grouping of ranges, e.g. [1-3]
	sorting=none,        % Numbers references in the order they are cited
	maxbibnames=6,       % List all authors up to 6
	minbibnames=3,       % If there are more than 6, list the first 3 followed by et al.
	doi=true,            % Include DOIs
	url=false,           % Hide URLs to keep it clean, unless there's no DOI
	isbn=true,           % Include ISBNs for books
	giveninits=true,     % Abbreviates first names to initials (like the examples)
	terseinits=false     % Keeps the periods after initials
]{biblatex}

\definecolor{authorcolor}{RGB}{90,90,90}
\definecolor{urlblue}{RGB}{20,20,190}
\renewcommand{\title}[1]{{\LARGE\sffamily\bfseries\selectfont\raggedright\textls{#1}\par\vskip.15in}}
\renewcommand{\author}[1]{{\raggedright\sffamily\bfseries\scshape\large\boldmath\color{authorcolor}#1\vskip1ex\par}}
\newcommand{\address}[1]{{\raggedright\small\itshape #1\par}}
\newcommand\authormark[1]{\textsuperscript{#1}}
\newcommand{\email}[1]{{\raggedright\footnotesize\itshape\color{urlblue}{#1}}\par}
\renewenvironment{abstract}{\vskip1pc\noindent\textbf{Abstract:}\space}{}

\begin{document}
\title{Sparse Spectral Imaging for Thickness Mapping of 3R-\ch{MoS2} on PDMS}

\author{Benjamin Laudert,\authormark{1,*} Fatemeh Abtahi,\authormark{1} Sarka Vavreckova,\authormark{1,2} Sebastian W. Schmitt,\authormark{2} and Falk Eilenberger\authormark{1,2,3}}

\address{
	\authormark{1}Institute of Applied Physics, Abbe Center of Photonics, Friedrich Schiller University Jena, Albert-Einstein-Str. 15, 07745 Jena, Germany\\
	\authormark{2}Fraunhofer-Institute for Applied Optics and Precision Engineering IOF, Albert-Einstein-Str. 7, 07745 Jena, Germany\\
	\authormark{3}Max Planck School of Photonics, Albert-Einstein-Str. 15, 07745 Jena, Germany
}

\email{\authormark{*}benjamin.laudert@uni-jena.de}

\begin{abstract}
We present a non-destructive, spatially resolved thickness characterization method for rhombohedral (3R) molybdenum disulfide (\ch{MoS2}) on polydimethylsiloxane (PDMS) substrates. Unlike broadband spectroscopic approaches, the proposed method reduces the measurement to a small number of discrete intensity images, enabling direct thickness mapping with a conventional microscope architecture and commercially available bandpass filters.
Our approach combines a systematic framework for selecting optimal discrete wavelength samples of the material's reflectance with a robust thickness retrieval algorithm based on a multivariate Gaussian probability model. 
By sampling the reflectance with just five strategically chosen near-infrared bandpass filters, we demonstrate thickness characterization up to \SI{691}{\nm} with a mean 95\% confidence-interval width of $\SI{8.3}{\nm}$. The method is adaptable to other van der Waals materials and conventional optical thin-film systems. It therefore provides a foundation for scalable, real-time thickness characterization in, e.g., dry-transfer fabrication workflows, where thickness screening remains a critical bottleneck for the production of van der Waals heterostructure devices.
\end{abstract}

%%%%%%%%%%%%%%%%%%%%%%%%%%  body  %%%%%%%%%%%%%%%%%%%%%%%%%%
\section{Introduction}
In recent years, there has been a growing interest in utilizing van der Waals materials in the optical thin-film regime for a multitude of photonic applications spanning the linear
\cite{muhammadOpticalBoundStates2021,
	zotevVanWaalsMaterials2023,
	munkhbatNanostructuredTransitionMetal2023,
	duCouplinginducedPerfectAbsorption2025},
nonlinear
\cite{wagonerSecondharmonicGenerationMolybdenum1998,
	zhaoAtomicallyPhasematchedSecondharmonic2016,
	shi3RMoS22017,
	xuCompactPhasematchedWaveguided2022,
	abdelwahabGiantSecondharmonicGeneration2022,
	hongTwistPhaseMatching2023,
	guoPolarizationEntanglementEnabled2024,
	qiStackingControlledGrowthRBN2024,
	peng3RstackedTransitionMetal2025,
	zografUltrathin3RMoS2Metasurfaces2025,
	seidtUltrafastAllopticalSwitching2025,
	zhuNonreciprocalLikeChiralSecondHarmonic2026},
and quantum
\cite{aharonovichQuantumEmittersHexagonal2022,
	guoUltrathinQuantumLight2023,
	weissflogTunableTransitionMetal2024,
	fengPolarizationentangledPhotonpairSource2024,
	tangQuasiphasematchingEnabledVan2024,
	trovatelloQuasiphasematchedDownconversionPeriodically2025,
	metuhIntegratedOnchipQuantum2025}
regimes.

One of the key parameters determining the optical response of such systems is the material thickness. However, deterministic growth of high-quality crystals with bottom-up techniques such as chemical vapor deposition remains challenging and requires multi-step processes \cite{moonHypotaxyWaferscaleSinglecrystal2025}, beyond the capabilities of most research groups.
Thus, mechanical exfoliation from a bulk crystal with subsequent transfer in a soft-polymer-based stamping process remains one of the most common and reliable techniques for obtaining high-quality single-crystal thin films.
However, as this is, fundamentally, a probabilistic process, obtaining the desired material thickness may require many attempts and screening of exfoliated flakes with respect to thickness is essential.

As such, it is highly inconvenient that conventional thickness characterization techniques, such as atomic force microscopy (AFM), surface profilometry, or vertical scanning interferometry (VSI), are perturbed by the soft, uneven surface of the transfer substrate to which the material is attached, fundamentally limiting their accuracy (see, for example, Supplementary Figure~S1). One is therefore forced to initially transfer vdW materials onto flat substrates for accurate thickness determination. This is especially problematic when a subsequent pick-up and transfer step is required (e.g. for integration in nanophotonic devices or heterostructures), which may lead to damage or destruction of the material. A direct characterization of vdW materials on polymers is therefore highly desirable.

Characterization techniques based on white light \cite{nguyenVisibilityHexagonalBoron2020a} or photoluminescence \cite{crimmannHighThroughputOpticalIdentification2025} microscopy have proven effective for few- and single-layer crystals. However, they are fundamentally limited in the maximum thickness that can be determined unambiguously. This limitation is especially severe for commonly used transition metal dichalcogenides (TMDs), such as molybdenum disulfide/diselenide (\ch{MoS2}/\ch{MoSe2}) and tungsten disulfide/diselenide (\ch{WS2}/\ch{WSe2}). These materials rapidly become optically opaque in the visible range with increasing thickness, producing responses that are difficult to distinguish from their bulk forms.

Recently, near-infrared extended reflection spectroscopy has emerged as a method for thickness characterization of vdW materials on PDMS substrates \cite{abtahiThicknessDependenceLinear2026}. The authors demonstrated that the thickness of rhombohedral \ch{MoS2} (3R-\ch{MoS2}) can be determined with \unit{\nm}-scale resolution. However, their approach relies on recording finely resolved broadband spectra, followed by least-squares optimization for film thickness retrieval. This approach is therefore unsuitable for the creation of complete thickness maps and requires a spectrometer as an additional measurement device.

To overcome these limitations, we introduce a sparse wavelength sampling technique that exploits the pronounced thickness-dependent reflectance of high-index thin films. The method enables rapid, spatially resolved thickness characterization directly on PDMS substrates using only a microscope and a small set of spectral filters. We demonstrate the approach using 3R-\ch{MoS2} as a case study. This material has emerged as a promising platform for nonlinear and quantum photonic applications because of its commercial availability, high refractive index, and substantial second-order nonlinear optical response \cite{wagonerSecondharmonicGenerationMolybdenum1998, xuCompactPhasematchedWaveguided2022}.

We first present a procedure for selecting suitable sets of bandpass filters. This procedure is based on a distinguishability metric defined over the target thickness interval and derived from the material’s reflectance response.
We then present reflectance measurements of mechanically exfoliated 3R-\ch{MoS2} flakes on PDMS substrates. These measurements are correlated with reference thickness values obtained by vertical scanning interferometry after transfer. This correlation allows us to refine the initially calculated relationship between thickness and reflectance. It also provides the residual statistics used to define a multivariate Gaussian model for the thickness retrieval algorithm.
Finally, we investigate the efficacy of the presented method for other vdW and conventional thin-film materials, covering a broad range of dielectric functions.

\section{Method and Results}\label{sec:results}

\subsection{Bandpass Filter Set Selection} \label{sec:BPF_selection}
\begin{figure}[htb]
	\centering
	\includegraphics{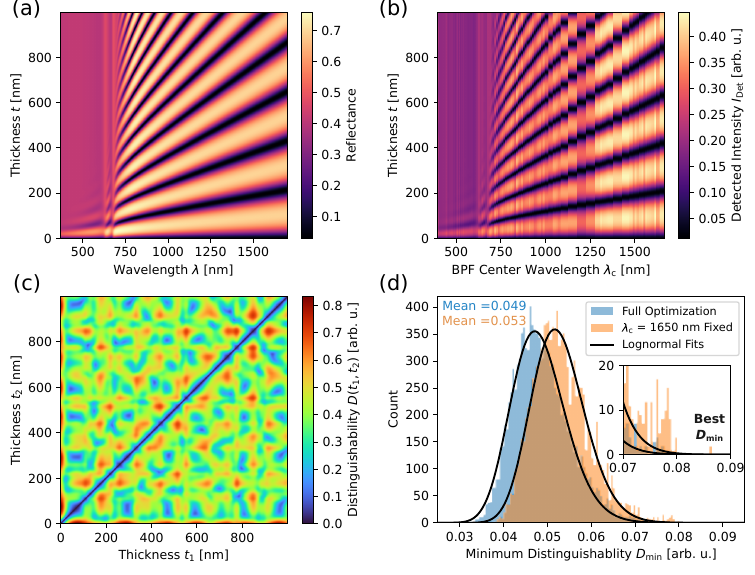}
	\caption{\textbf{Optical modeling and bandpass filter set optimization.}
	\textbf{(a)} Calculated reflectance $R$ for a 3R-\ch{MoS2} thin film on PDMS of thickness $t$ at wavelength $\lambda$.
    \textbf{(b)} Calculated thickness dependent intensity $I_\mathrm{Det}$ for imaging $R$ through each considered bandpass filter.
    \textbf{(c)} Distinguishability matrix for the optimal bandpass filter set $\Lambda = \{ \SI{920}{\nm}, \SI{1050}{\nm}, \SI{1350}{\nm}, \SI{1450}{\nm}, \SI{1650}{\nm} \}$.
    \textbf{(d)} Histograms of minimum distinguishability values $D_\mathrm{min}$ resulting from the simulated annealing optimization processes and fitted lognormal distribution curves.}
	\label{fig:fig1_BPF_opt}
    \centering
\end{figure}

To identify a suitable combination of bandpass filters, we first calculate the expected reflectance $R$ of a 3R-\ch{MoS2} thin film on a PDMS substrate. The reflectance is calculated as a function of thickness $t$ and wavelength $\lambda$ using the optical transfer matrix method described in Section~\ref{sec:meth:tmm}. The calculation assumes normally incident light, and is based purely on the in-plane refractive index of 3R-\ch{MoS2} from Ref.~\cite{xuCompactPhasematchedWaveguided2022} and the isotropic refractive index of PDMS from Ref.~\cite{guptaMechanotunableSurfaceLattice2019}. The result is shown in Figure~\ref{fig:fig1_BPF_opt}(a) for our thickness range of interest ($\SI{0}{\nm} \leq t \leq \SI{1000}{\nm}$).

In Figure~\ref{fig:fig1_BPF_opt}(a), we can clearly see how, beyond the cutoff of absorption at $\lambda \approx \SI{700}{\nm}$, $R$ oscillates with increasing $t$ at any given value of $\lambda$, which is the fundamental effect underpinning the thickness retrieval algorithm, which will be presented in Section~\ref{sec:thick_det}.

To enable unambiguous thickness determination beyond half a period of the reflectance oscillations, sampling at multiple wavelengths is required. A spatially resolved implementation can be achieved using various multispectral or hyperspectral imaging techniques. These techniques involve different trade-offs between measurement time, spatial resolution, spectral resolution, spectral bandwidth, and computational complexity \cite{plazaRecentAdvancesTechniques2009, arceCompressiveCodedAperture2014, parkMicrosphereassistedHyperspectralImaging2024}.
In this work, we use a simple spectral filtering approach based on commercially available bandpass filters \cite{thorlabsincSpectralFilters2026}. This enables robust and rapid image acquisition without intrinsic loss of resolution and can, in principle, be performed in a fully parallel fashion, if sufficiently few wavelength samples are required.

We next calculate the total detected intensity for each bandpass filter. For a 3R-\ch{MoS2} thin film of thickness $t$, this intensity is obtained by integrating the product of the reflection spectrum $R(\lambda,t)$ and the filter transmission spectrum $T_{\lambda_\mathrm{c}}(\lambda)$ over wavelength:
\begin{align}
	I_\mathrm{Det}^{\lambda_\mathrm{c}}(t) = \int T_{\lambda_\mathrm{c}}(\lambda) R(\lambda, t) \mathrm{d}\lambda
\end{align} 
The resulting discretized spectral response is displayed in Figure~\ref{fig:fig1_BPF_opt}(b). Note that the continuous spectral response is approximated reasonably well, due to the narrow $\approx \SI{10}{\nm}$ filter transmission window and the large total number of considered filters $n = 100$.

To serve as the foundation for our bandpass filter selection, we then define the measure of distinguishability $D$ based on any set of $k$ bandpass filters $\Lambda = \left\{ \lambda_{\mathrm{c}_1}, ..., \lambda_{\mathrm{c}_k} \right\}$, for any two given thickness values $t_1$ and $t_2$ which are to be discerned based on a measurement of $I_\mathrm{Det}$:
\begin{align}
	D(t_1, t_2, \Lambda) = \left( \sum_{\lambda_\mathrm{c} \in \Lambda } \left( I_\mathrm{Det}^{\lambda_{\mathrm{c}}}(t_1) - I_\mathrm{Det}^{\lambda_{\mathrm{c}}}(t_2)\right)^2 \right)^\frac{1}{2},
\end{align}
which can be interpreted as the Euclidean distance between two points in a $k$-dimensional space, both of which are constrained to the same trajectory given by the $\lambda_\mathrm{c}$ components of $I_\mathrm{Det}^{\lambda_\mathrm{c}}(t)$. Note that for $t_1 = t_2$, the distinguishability becomes zero, while for unequal values $D$ is likely to increase with rising dimensionality $k$. In Figure~\ref{fig:fig1_BPF_opt}(c), we have displayed $D$ as a matrix of thickness values $t_1$ and $t_2$ on a \SI{1}{\nm} grid for a set of bandpass filters particularly well suited for thickness determination, which we obtained from the optimization approach described in the following.

Based on $D$ and our target thickness interval $T = [\SI{0}{\nm}, \SI{1000}{\nm}]$, we define the optimization function
\begin{align}
	D_\mathrm{min} (\Lambda) = \min_{(t_1, t_2)} \left\{ D(t_1, t_2, \Lambda):  t_1, t_2 \in T \ \mathrm{and} \ |t_1 - t_2| \geq \SI{10}{\nm} \right\}
\end{align}
which yields the minimum distinguishability for all pairs of thicknesses in the target thickness interval, excluding the trivial $D \approx 0$ region around $t_1 = t_2$. Therefore, we seek to maximize $D_\mathrm{min}$ to avoid errors due to the confusion of two thickness values with similar reflectance samples.
Our choice of optimization target is not fundamental. A different function based on $D(t_1,t_2)$ could be chosen to balance measurement precision against the risk of confusing different thickness values. If measurement uncertainties are known in advance, they could also be included explicitly to set desired probability limits for misclassification.

Here, we consider the case of $k = 5$ bandpass filters, to allow for practical operation in a standard 6-position filter wheel, while leaving one slot unoccupied for unfiltered acquisition.
Even for this modest number, there are $\approx \num{9.2e7}$ possible $\Lambda$, as they form a multiset under evaluation of $D$. A direct, brute-force search for the optimal $\Lambda$ is therefore not computationally practical for us. Furthermore, gradient-descent-based optimization methods cannot be employed due to the discrete nature of the bandpass filters. However, as the total number of considered bandpass filters is fairly large, adjacency effects exist when the bandpass filters are ordered based on $\lambda_\mathrm{c}$. Correspondingly, we decided to employ gradient-free simulated annealing to search for maxima in $D_\mathrm{min}$.

The resulting distribution of optimization results for a total of $N\approx 11000$ attempts is shown in Figure~\ref{fig:fig1_BPF_opt}(d). In addition to a fully unconstrained optimization, in which all $k=5$ filters are varied freely, we also tested a constrained strategy. In this case, one filter was fixed at the largest center wavelength $\lambda_\mathrm{c} = \SI{1650}{\nm}$. This filter has the longest reflectance oscillation period with respect to thickness $t$, and therefore provides the longest interval over which the reflectance can be inverted unambiguously using a single wavelength.

For both optimization strategies, the resulting distribution of $D_\mathrm{min}$ approximates a lognormal distribution (black curves). Additionally, we observe a significant advantage for the constrained strategy, as the mean of its distribution $\overline{D}_\mathrm{min} = 0.053$ is approximately $8\%$ greater than that of the full optimization ($\overline{D}_\mathrm{min} = 0.049$), while their standard deviations are almost equal at $\sigma_{D_\mathrm{min}} \approx 0.065$. The best combination of bandpass filters we found across both optimization strategies was
$\Lambda = \{ \SI{920}{\nm}, \allowbreak \SI{1050}{\nm}, \allowbreak \SI{1350}{\nm}, \allowbreak \SI{1450}{\nm}, \allowbreak \SI{1650}{\nm} \}$
which has a corresponding minimum separation of $D_\mathrm{min} = 0.086$.
The corresponding separation matrix is displayed in Figure~\ref{fig:fig1_BPF_opt}(c) and has its minima at $(t_1, t_2) = (\SI{932}{\nm}, \SI{744}{\nm})$ and vice versa.

\subsection{Sample Characterization}
\begin{figure}[htb]
	\centering
	\includegraphics{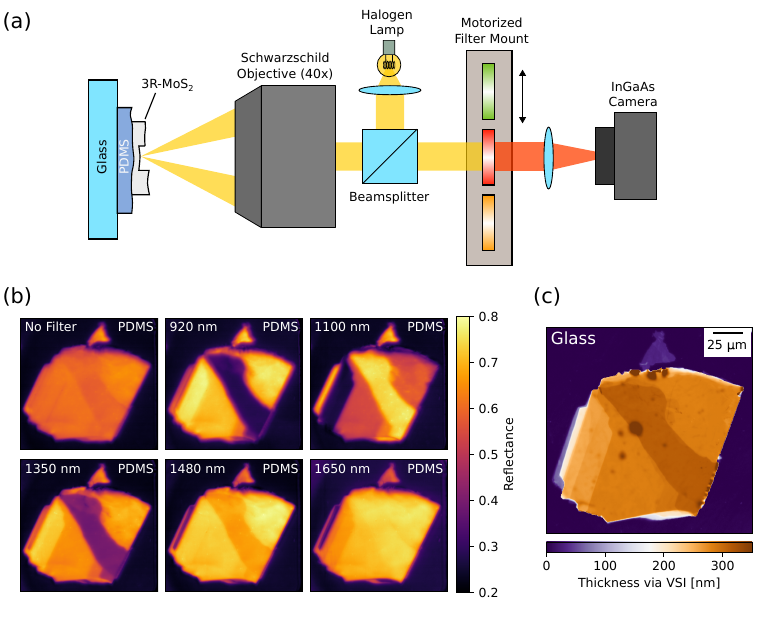}
	\caption{\textbf{Multispectral imaging microscope and thickness validation.}
		\textbf{(a)} Multispectral imaging microscope used for thickness mapping of 3R-\ch{MoS2} on PDMS. The broadband NIR radiation from a halogen lamp is reflected off the sample, spectrally filtered through bandpass filters and imaged with an InGaAs camera.
		\textbf{(b)} Recorded NIR images of a 3R-\ch{MoS2} flake on a PDMS substrate. The sections with different thicknesses each show a unique response for each filter due to thin-film interference.
		\textbf{(c)} Material thickness of the sample in (b) measured by vertical scanning interferometry (VSI) after transfer onto a glass substrate.}
	\label{fig:fig2_setup_and_example_measurement}
\end{figure}

To test the method, we prepared multiple samples of 3R-\ch{MoS2} flakes on PDMS as described in the Methods section. For each sample, we recorded reflection images with the bandpass filter set 
$\Lambda = \{ \SI{920}{\nm}, \allowbreak \SI{1100}{\nm}, \allowbreak \SI{1350}{\nm}, \allowbreak \SI{1480}{\nm}, \allowbreak \SI{1650}{\nm} \}$
using the microscopy setup depicted in Figure~\ref{fig:fig2_setup_and_example_measurement}(a).
Note that, due to availability, we used a slightly less optimal set of filters with a slightly lower $D_\mathrm{min}$ value of 0.078.
A halogen lamp was used as a broadband, near-infrared (NIR) illumination source, which was focused onto the sample with a Schwarzschild objective.
The light reflected off the sample was imaged with an indium gallium arsenide (InGaAs)-based NIR camera and referenced to a silver mirror, as described in the Methods section.

After the multispectral image acquisition, each sample was transferred onto a flat fused silica glass substrate and characterized via vertical scanning interferometry (VSI), which we used as a reference thickness characterization method.

In Figure~\ref{fig:fig2_setup_and_example_measurement}(b) we have displayed the filtered reflectance images of one of our studied samples, alongside an image recorded without any spectral filtering. The reference thickness map, obtained via VSI after transfer, is displayed in Figure \ref{fig:fig2_setup_and_example_measurement}(c) and shows that the flake has several distinct sections with different thicknesses.
Due to the broadband spectral sensitivity of the camera sensor, the contrast between the different sections is fairly low in the unfiltered image. In the spectrally filtered images, however, the sections can be distinguished much more clearly, as strong reflectance modulations in the range of $R = 0.20$ and $R=0.75$ are observed with respect to the bandpass filter center wavelength.

\subsection{Reflectance Model Refinement}
\begin{figure}[htbp]
	\centering
	\includegraphics{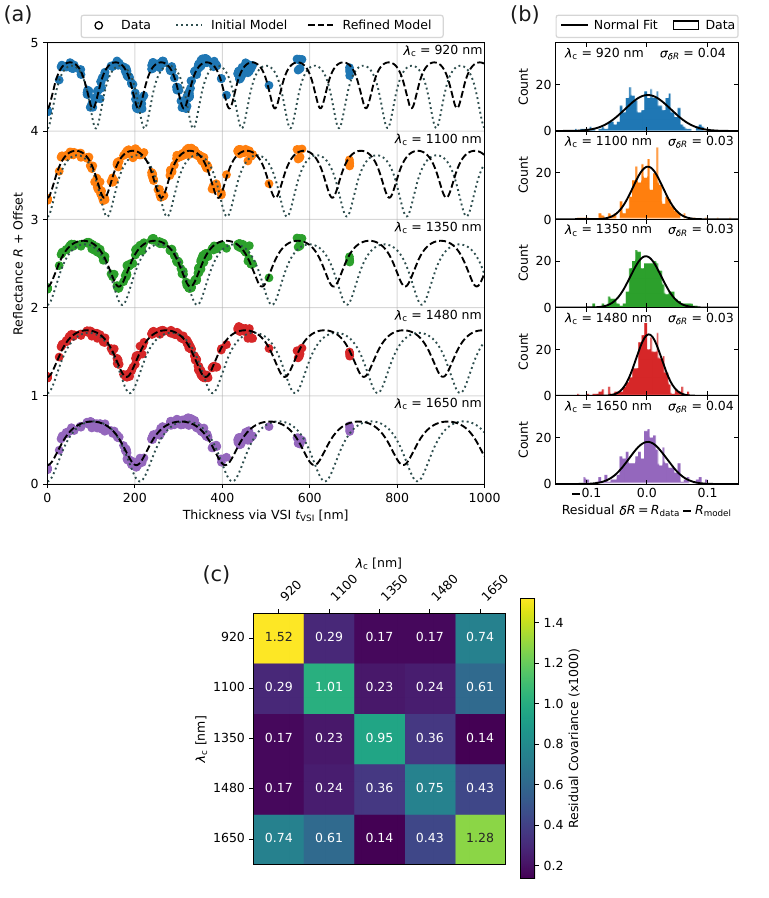}
	\caption{\textbf{Experimental validation and refinement of the reflectance model.}
		\textbf{(a)} Measured reflectance values for each bandpass filter center wavelength $\lambda_\mathrm{c}$ as a function of thickness as determined through VSI after transfer. Additionally, the initially calculated and refined models are displayed.
		\textbf{(b)} Histogram of the residuals between the measured data and the refined model and fitted normal distribution curves.
		\textbf{(c)} Covariance matrix of residuals.
	}
	\label{fig:fig3_model_refinement}
\end{figure}

In order to exploit the interference effect for thickness characterization, we model the expected reflectance as a function of thickness for each bandpass filter channel, taking into account the distribution of incidence angles of the Schwarzschild objective as described in Section~\ref{sec:ss_obj_modeling}. The resulting reflectance curves form our initial model and are displayed as the dotted lines in Figure~\ref{fig:fig3_model_refinement}(a). Initially, we attempted to determine the material thickness based on this model, however, in doing so, we observed a severe mismatch between the thickness values obtained via VSI and those obtained via the reflectance-based method.

We therefore performed a preliminary refinement procedure for the reflectance model, based on a dataset of matching pairs of reflectance and thickness values, obtained via manual selection as detailed in Section~\ref{sec:meth:msi}. Note that we carefully avoided any edges or film imperfections to exclusively probe the thin-film response.

In Figure~\ref{fig:fig3_model_refinement}(a), we have displayed the reflectance values of the refinement dataset against their corresponding thickness values obtained through VSI.
The data qualitatively agrees with the initial model as regular oscillations with respect to thickness are observed. However, quantitatively, two deviations are shared systematically across all wavelength channels:
Compared to the initial model, the oscillation amplitudes of the experimentally observed reflectance values are reduced, which is mainly caused by an increase of the lower limit of reflectance modulation from $\min(R_\mathrm{model}) = 0.03$ to $\min(R_\mathrm{data}) \approx 0.20$, while the upper limits remain similar at $\max(R_\mathrm{model}) = 0.74$ and $\max(R_\mathrm{data}) \approx 0.75$. We attribute this to reflections on the backside of the PDMS substrate, which can propagate back through the flake when transmittance is high, i.e., reflectance is low.
Additionally, we observe a shorter oscillation period for the reflectance data compared to the model, leading to a gradual dephasing effect as the flake thickness increases. This is most likely caused by an underestimation of the refractive index of our 3R-\ch{MoS2} material, as we used the refractive index data provided in Ref.~\cite{xuCompactPhasematchedWaveguided2022} for our calculations.

To account for these effects, we introduce three correction parameters $\alpha$, $\beta$, and $\gamma$ for each $\lambda_\mathrm{c}$ channel, such that the refined reflectance is given by:
\begin{align}
	R'_{\lambda_\mathrm{c}}(t) = \alpha + \beta R_{\lambda_\mathrm{c}}(\gamma t)
\end{align}
Here, $\alpha$ and $\beta$ correct the offset and reduction of the oscillation amplitude caused by the back reflections, while $\gamma$ corrects the dephasing resulting from the refractive index mismatch.

These parameters are optimized for each channel via least-squares minimization. The resulting values for $\alpha$, $\beta$, and $\gamma$ are displayed in Table \ref{tab:refinement_params} and the refined reflectance curves are displayed in Figure~\ref{fig:fig3_model_refinement}(a) as dashed, black lines.

In Figure~\ref{fig:fig3_model_refinement}(b), we have displayed histograms of the residuals of the refined model to the reflectance data for each bandpass filter channel, which show that the residuals are distributed reasonably close to normal distributions with standard deviations ranging from $\sigma_{\delta R} = 0.027$ to $\sigma_{\delta R} = 0.038$.

Furthermore, as shown in Figure~\ref{fig:fig3_model_refinement}(c), the calculation of the overall residual covariances across the five $\lambda_c$ channels shows a strong positive correlation, indicating that, on average, residuals deviate systematically rather than independently, which implies that systematic deviations from the reflectance model are a substantial contributor to measurement error.
Pinpointing the underlying mechanisms of these correlations is challenging and beyond the scope of this work, however, a likely contribution arises from wavy deformations of the PDMS substrates and attached 3R-\ch{MoS2} flakes, which perturb the optical response of the thin-film systems. Nonetheless, even without precise knowledge of their origin, these statistical correlations can be exploited for improved robustness in the thickness retrieval process, as we will show in the following section.

\begin{table}[htbp]
	\centering
	\caption{Optimized refinement parameters for each bandpass filter channel $\lambda_c$.}
	\begin{tabular}{cccccc}
		\toprule
		Parameter & \SI{920}{\nm} & \SI{1100}{\nm} & \SI{1350}{\nm} & \SI{1480}{\nm} & \SI{1650}{\nm} \\
		\midrule
		$\alpha$  & 0.23          & 0.22           & 0.21           & 0.19           & 0.19           \\
		$\beta$   & 0.74          & 0.77           & 0.76           & 0.77           & 0.73           \\
		$\gamma$  & 1.07          & 1.05           & 1.05           & 1.04           & 1.04           \\
		\bottomrule
	\end{tabular}
	\label{tab:refinement_params}
\end{table}

\subsection{Thickness Retrieval} \label{sec:thick_det}
Every data point in a multispectral image represents a measurement of multiple reflectance values that can be arranged into a vector $\vec R_\mathrm{meas}$.
Based on the refined reflectance model $\vec{R}_\mathrm{model}$, a vector of residuals can be calculated for each potential thickness $t$ by subtracting the corresponding modeled vector of reflectance values:
\begin{align}
	\delta \vec R (t) =  \vec R_\mathrm{meas} - \vec R_\mathrm{model}(t)
\end{align}
Analyzing this vector of residuals with respect to the multivariate Gaussian (global) probability density function (PDF) of all residuals \cite{deisenrothMathematicsMachineLearning2020}:
\begin{align}
	\rho_\mathrm{G}(\delta \vec{R}) = \frac{1}{\sqrt{(2\pi)^k \det(\Sigma)}} \exp \left( -\frac{1}{2} \delta \vec{R}^T \Sigma^{-1} \delta\vec{R} \right), \label{eq:global_PDF}
\end{align}
where $\Sigma$ is the covariance matrix of residuals shown in Figure~\ref{fig:fig3_model_refinement}(c), forms the foundation of our thickness retrieval algorithm.

\begin{figure}[htbp]
	\centering
	\includegraphics[width=\textwidth]{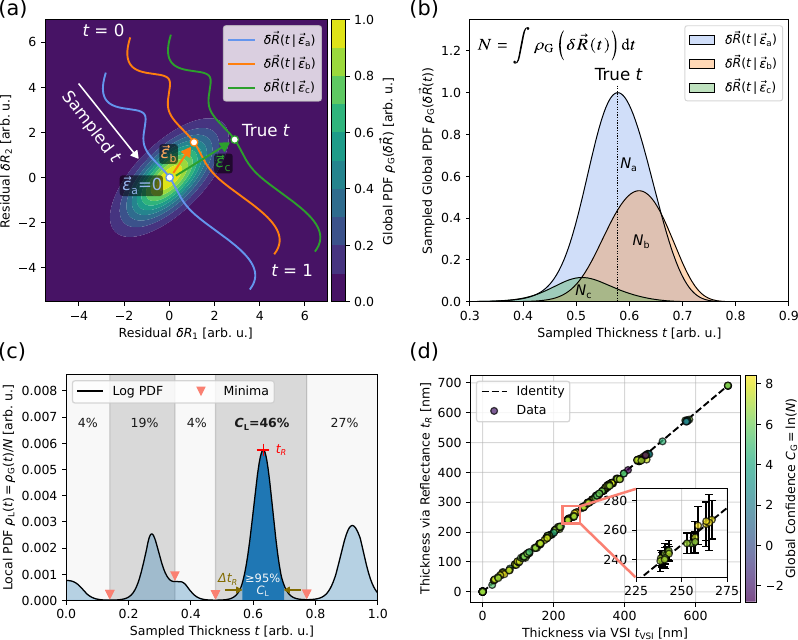}
	\caption{\textbf{Thickness retrieval algorithm and statistical confidence framework.}
	    \textbf{(a)} Illustration of a global probability density function (PDF) of residuals $\rho_\mathrm{G}(\delta R_1, \delta R_2)$. Increasing reflectance measurement error $\vec{\varepsilon}$ shifts the sampled trajectory of residuals away from the maximum value at the origin.
		\textbf{(b)} Values of $\rho_\mathrm{G}$ sampled by the trajectories displayed in (a). With increasing measurement error $\vec{\varepsilon}$, the integral of sampled probabilities $N$ decreases.
		\textbf{(c)} Illustration of the segmentation and thickness classification steps for the local PDF $\rho_\mathrm{L}(t)$ from which the local confidence $C_\mathrm{L}$, optimal thickness $t_R$ and precision $\Delta t_R$ are obtained.
		\textbf{(d)} Thickness values $t_R$ obtained by performing the thickness retrieval algorithm on the model refinement dataset, plotted against their corresponding values obtained via VSI $t_\mathrm{VSI}$. The errorbars shown in the inset are $\Delta t_R$, while the color scale encodes the global confidence metric $C_\mathrm{G} = \ln{N}$.}
	\label{fig:fig4_model_thickness_evaluation}
\end{figure}

Initially, $\rho_\mathrm{G}$ is sampled with respect to $t$. This is illustrated for a synthetic example in Figure~\ref{fig:fig4_model_thickness_evaluation}(a) and (b) for varying levels of measurement error $\vec \varepsilon$ for the case of $k=2$ bandpass filter channels.
In the absence of error $\vec{\varepsilon}_\mathrm{a} = 0$, the trajectory of residuals traced by $\delta \vec{R}(t | \varepsilon_\mathrm{a})$ intersects the maximum of the PDF at the origin at the true underlying value of $t$.
A non-zero measurement error, as illustrated for $\vec \varepsilon_\mathrm{b}$ and $\vec \varepsilon_\mathrm{c}$ in Figure~\ref{fig:fig4_model_thickness_evaluation}(a), shifts this trajectory away from the origin, which, as displayed in Figure~\ref{fig:fig4_model_thickness_evaluation}(b), may offset the maximum of the sampled curve away from the true underlying thickness value.

Additionally, as the sampled trajectory is shifted away from the maximum of the residual PDF, the total sampled amount of $\rho_\mathrm{G}$
\begin{align}
	N = \int_{t_\mathrm{min}}^{t_\mathrm{max}} \rho_\mathrm{G}\left( \delta \vec{R}(t) \right) \mathrm{d}t
\end{align}
is reduced. Although $N$ does not represent an absolute probability, it provides a measure of quality for the probed combination of residuals.
As we have found $N$ to vary over multiple orders of magnitude experimentally, we use its natural logarithm as a global confidence metric
\begin{align}
    C_\mathrm{G} = \ln(N).
\end{align}

Furthermore, any sampled trajectory $\rho_\mathrm{G}(\delta \vec{R}(t))$ can be converted into a local PDF of potential thickness values by normalizing to $N$:
\begin{align}
	\rho_\mathrm{L}(t) = \frac{1}{N} \rho_\mathrm{G}\left(\delta \vec{R}(t)\right).
\end{align}
The local PDF is then segmented at its minima to generate a series of thickness intervals, each of which has a certain probability of containing the true thickness. This is illustrated in Figure~\ref{fig:fig4_model_thickness_evaluation}(c) for a synthetic $\rho_\mathrm{L}(t)$, possessing many local maxima. Note that this segmentation is performed directly on the logarithmic PDF, i.e.~the argument inside $\exp$ in Equation~\eqref{eq:global_PDF}, for improved numerical stability.

To obtain a single thickness value, we choose the segment with the highest probability and select the maximum of $\rho_\mathrm{L}$ inside the PDF segment as the most likely thickness estimate $t_R$, while the segment probability is taken as a second quality metric (local confidence $C_\mathrm{L}$). Finally, to estimate the precision of the obtained thickness value, we integrate the local PDF around $t_R$ until $95\%$ of $C_\mathrm{L}$ is reached, with the width of the resulting thickness interval yielding the precision $\Delta t_R$.

This hierarchical approach guards against several types of misclassification errors. In the case of large noise and total model breakdown, the global confidence $C_\mathrm{G}$ is severely reduced.
If the trajectory of residuals is displaced by noise in such a way that two or more segments of different thicknesses become likely, $C_\mathrm{L}$ decreases.
For a local PDF with a broad peak around $t_R$, the precision decreases, as $\Delta t_R$ increases.
Note, however, that this assumes that the true film thickness falls inside the sampled interval. As we have chosen a sufficiently large $t_\mathrm{max} = \SI{1000}{\nm}$, this can easily be verified with e.g.~standard white light microscopy using the depth-of-field effect of a high NA objective.

In Figure~\ref{fig:fig4_model_thickness_evaluation}(d) we have applied the thickness retrieval algorithm to the refinement dataset and see good agreement with the reference thickness obtained via VSI. For $m = 287$ of the total $n = 293$ data points ($m/n = 98\%$), we find that $t_R$ agrees with $t_\mathrm{VSI}$ within the window of precision, for which we find a mean value of $\overline{\Delta t_R} = \SI{8.3}{\nm}$ and a maximum value of $\max(\Delta t_R) = \SI{23}{\nm}$ across the entire dataset.
Similarly, we find that $m = 286$ of the sampled data points fulfill $C_\mathrm{L} > 0.95$, with the remaining data points having varying values of $C_\mathrm{L}$ ranging from $0.61$ to $0.85$.
For the global confidence, we find that $95\%$ of values lie above $C_\mathrm{G} = 4.0$, closer to the maximum of the obtained values $\max( C_\mathrm{G} ) = 8.4$ than the minimum $\min(C_\mathrm{G}) = -2.8$.

\begin{figure}[htb]
	\centering
	\includegraphics{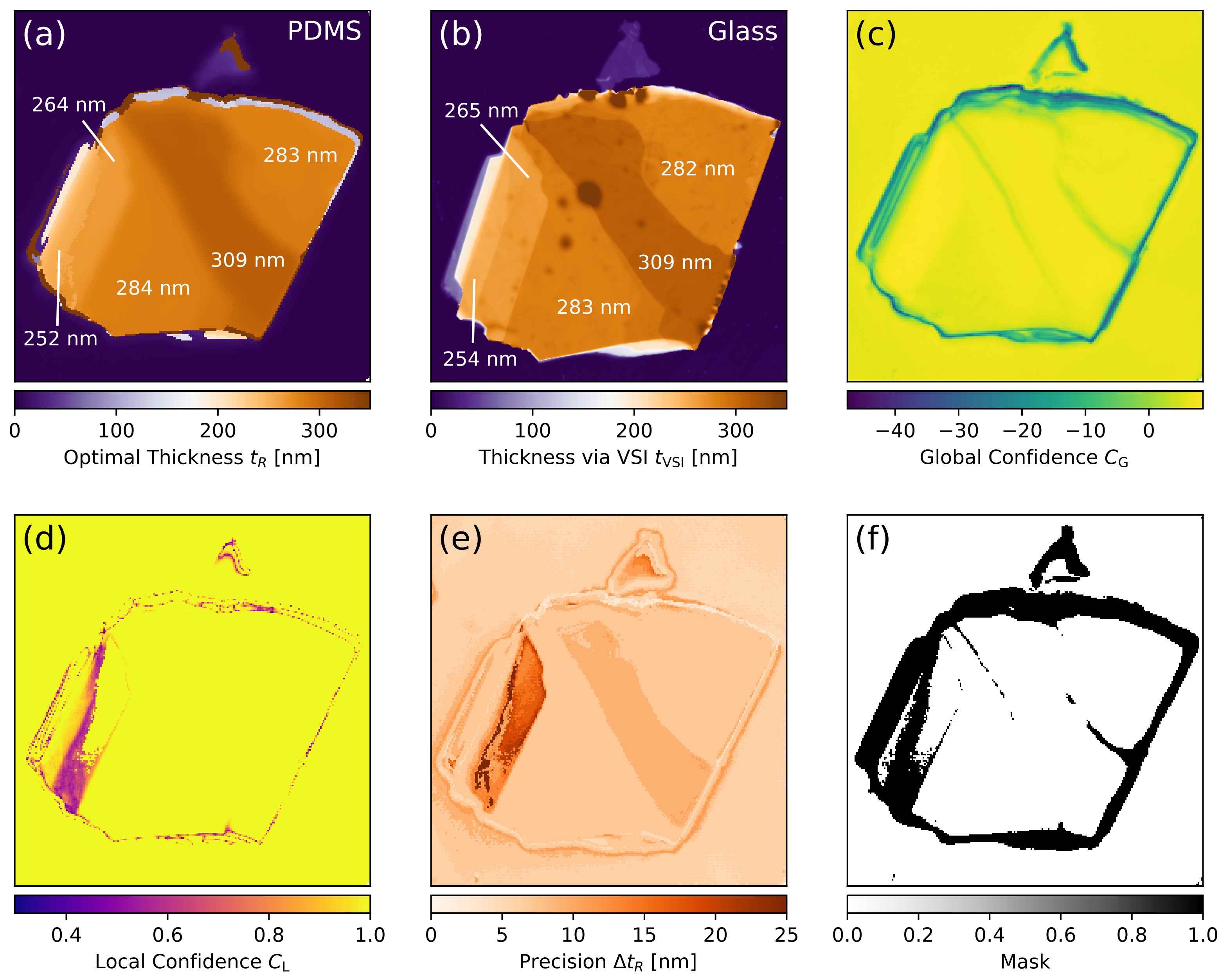}
	\caption{\textbf{Thickness mapping and confidence analysis.}
		(a) Thickness map of the 3R-\ch{MoS2} flake on PDMS displayed in Figure~\ref{fig:fig2_setup_and_example_measurement}, obtained with the reflectance-based method.
		(b) Material thickness as determined through VSI after transfer of the flake onto a glass substrate.
		(c)-(e) Maps of global confidence $C_\mathrm{G}$, local confidence $C_\mathrm{L}$, and precision $\Delta t_R$ derived from the reflectance-based thickness determination method. (f) Combined mask based on heuristic limits for $C_\mathrm{G}$, $C_\mathrm{L}$, and $\Delta t_R$ which indicates potentially incorrect thickness values.
	}
	\label{fig:fig5_thickness_image}
\end{figure}

We next evaluate the method for spatially resolved thickness characterization. In this case, deviations from the ideal thin-film response are expected. We therefore apply the thickness retrieval procedure to each pixel of the reflectance images shown in Figure~\ref{fig:fig2_setup_and_example_measurement}(b).
In Figure~\ref{fig:fig5_thickness_image}(a) the thickness of the 3R-\ch{MoS2} flake, as determined through the reflectance-based method on PDMS before transfer is displayed.
Figure~\ref{fig:fig5_thickness_image}(b) shows the reference thickness map obtained by VSI after transfer to glass. Compared with this reference, the reflectance-based thickness map shows excellent agreement for the flat, pristine regions of the flake, with deviations of only a few nanometers. Around the flake edges and in narrow regions on the left-hand side and bottom of the flake, some thickness values are misclassified. We attribute these errors to the intermixing of reflectance values caused by the limited spatial resolution of the microscope.

Accordingly, at least one of the three confidence metrics, shown in Figure~\ref{fig:fig5_thickness_image}(c)-(e), is reduced in those parts of the image. By applying the heuristic limits of $C_\mathrm{G} \geq 4$, $C_\mathrm{L} \geq 0.95$, and $\Delta t_R \leq \SI{25}{\nm}$, based on the previous experimental observations, we generate the combined binary mask in Figure~\ref{fig:fig5_thickness_image}(f), which highlights the corresponding regions of the image likely to contain misclassified thickness values.

Furthermore, as we observed a model-intrinsic dependence of $C_\mathrm{G}$, $C_\mathrm{L}$, and $\Delta t_R$ on the probed thickness, as shown in Figures~S2 and S3, which could potentially make the statistical model overly specific to the refinement dataset, we also applied the thickness retrieval algorithm to a separate set of samples. In doing so, we observed good agreement between $t_\mathrm{VSI}$ and $t_R$ within the window of precision given by $\Delta t_R$ in the vast majority of cases (Figure~S4). This indicates that strong overspecialization did not occur and that the refinement dataset is a representative sample of 3R-\ch{MoS2} thin films on PDMS in the studied thickness range.

\subsection{Applicability to Other Materials}
In principle, the thickness characterization methodology presented here can be applied to any material which has a refractive index contrast to the underlying substrate and is sufficiently transparent in the spectral range of detection. However, a change in the refractive index of the examined material leads to a change in the oscillation amplitude and frequency with respect to thickness for any given wavelength.

To investigate how variations in refractive index affect the method for vdW materials beyond 3R-\ch{MoS2}, we apply the bandpass filter selection method (detailed in Section~\ref{sec:BPF_selection}) using refractive index data for hBN \cite{grudininHexagonalBoronNitride2023}, \ch{MoSe2} \cite{munkhbatOpticalConstantsSeveral2022}, and \ch{MoTe2} \cite{munkhbatOpticalConstantsSeveral2022}. Notably, hBN exhibits a refractive index substantially lower than that of 3R-\ch{MoS2}, while \ch{MoSe2} displays a dielectric response comparable to, but slightly stronger than, that of 3R-\ch{MoS2}. In contrast, \ch{MoTe2} possesses a significantly higher refractive index, with an absorption cutoff in the near-infrared at $\lambda \approx \SI{1350}{\nm}$. Together, these materials span a broad range of optical properties. They therefore serve as representative examples of vdW materials that are potentially compatible with this technique.

Table \ref{tab:other_vdW_bpf_select_res} displays the results obtained from the bandpass filter selection scheme for these vdW materials, as well as the previously obtained values for 3R-\ch{MoS2}. Note that a comprehensive overview of the optimization results is presented in Figure~S5.
We observe a substantial dependence of the obtained best (highest) value of $D_\mathrm{min}$ on the studied material, with hBN showing the lowest minimum distinguishability value of $D_\mathrm{min}^\mathrm{hBN} = 0.045$ among the tested vdW materials. hBN also has the lowest refractive index among the materials considered here, with $n\approx2.2$. We therefore attribute the reduced $D_\mathrm{min}$ to the smaller reflectance oscillation amplitude with respect to film thickness. This smaller amplitude results from the lower refractive index contrast with the PDMS substrate, which has $n\approx1.4$.
This limitation is partially compensated by the reduced oscillation frequency. As a result, the distinguishability matrix varies more slowly and contains fewer local minima, as shown in Figure~S5(d). In addition, the wide transparency window of hBN enables sampling across the full set of considered bandpass filters.

\newcommand{\cellmostwo}{\makecell{920\\ 1050\\ 1350\\ 1450 \\ 1650}}
\newcommand{\cellhbn}{\makecell{580\\ 610\\ 780\\ 1320\\ 1650}}
\newcommand{\cellmosetwo}{\makecell{990\\ 1030\\ 1300\\ 1500\\ 1650}}
\newcommand{\cellmotetwo}{\makecell{1250\\ 1300\\  1480\\ 1500\\ 1650}}
\newcommand{\cellGaP}{\makecell{730\\ 920\\  1030\\ 1450\\ 1650}}
\newcommand{\cellSi}{\makecell{850\\ 1030\\  1330\\ 1450\\ 1650}}
\newcommand{\cellSiO}{\makecell{400\\ 420\\ 450\\ 510\\ 950}}
\begin{table}[htbp]
    \centering
    \caption{Bandpass filter optimization results for hBN, \ch{MoS2}, \ch{MoSe2}, and \ch{MoTe2} on PDMS, as well as conventional thin-film systems of \ch{GaP} and \ch{Si} on insulator (\ch{SiO2}), as well as oxidized silicon (\ch{SiO2} on \ch{Si}). Additionally, the absorption cutoff wavelength, above which $k \leq 0.1$, is listed for each material.}
    \begin{tabular}{cccccccc}
    \toprule
     & \makecell{3R-\ch{MoS2}\\on PDMS}
     & \makecell{hBN on\\PDMS }
     & \makecell{\ch{MoSe2} on\\PDMS}
     & \makecell{\ch{MoTe2} on\\PDMS}
     & \makecell{\ch{GaP}\\on \ch{SiO2}}
     & \makecell{\ch{Si} on\\\ch{SiO2}}
     & \makecell{\ch{SiO2}\\on \ch{Si}} \\
    \midrule
    \makecell{Absorption\\Cutoff $\lambda$}
    &  \SI{720}{\nm}
    & -
    & \SI{858}{\nm}
    & \SI{1350}{\nm}
    & \SI{476}{\nm}
    & \SI{450}{\nm}
    &  - \\
    Best $D_\mathrm{min}$ & 0.086   & 0.045   & 0.077   & 0.051 & 0.076 & 0.089 & 0.040 \\[0.2cm]
    \makecell{Optimal $\lambda_\mathrm{c}$\\ {[nm]}}
    & \cellmostwo
    & \cellhbn
    & \cellmosetwo
    & \cellmotetwo
    & \cellGaP
    & \cellSi
    & \cellSiO \\
    \bottomrule
    \end{tabular}
    \label{tab:other_vdW_bpf_select_res}
\end{table}

For \ch{MoTe2}, the opposite is the case. Due to the material's large refractive index of $n \approx 4.8$ inside the window of transparency, the refractive index contrast to the substrate and the reflectance oscillation amplitudes are large, while numerous off-diagonal minima are present in the distinguishability matrix (see Figure~S5(f)). Furthermore, two of the optimal $\lambda_\mathrm{c}$ lie below the absorption cutoff at $\lambda = \SI{1350}{\nm}$, which indicates that sampling a broader range of frequencies is advantageous for such a broadly absorbing material, even if the oscillation amplitude tapers off towards the upper limit of the target thickness interval, as is apparent in Figure~S7.

Unsurprisingly, as \ch{MoSe2} is optically very similar to 3R-\ch{MoS2}, we obtain similar values of $D_\mathrm{min}$, with $D_\mathrm{min}^\mathrm{MoSe2} = 0.077$ being only slightly lower than $D_\mathrm{min}^\mathrm{MoS2} = 0.086$, and similar sets of optimal bandpass filters, both starting at around $\lambda_\mathrm{c}\approx \SI{950}{\nm}$.

Furthermore, we investigate the thickness characterization method for optical thin-film systems based on conventional materials. Specifically, we perform the bandpass filter selection scheme for gallium phosphide (GaP) and silicon (Si) on insulator (\ch{SiO2}) substrates, as well as thermally oxidized silicon (\ch{SiO2} on a Si substrate), using the refractive index data from Refs.~\cite{khmelevskaiaDirectlyGrownCrystalline2021, frantaTemperaturedependentDispersionModel2017, malitsonInterspecimenComparisonRefractive1965}. 
Note that Si and GaP on insulator both represent systems optically similar to the previously examined high-index vdW materials on low-index PDMS substrates, while \ch{SiO2} on Si represents the inverse case.

For the conventional optical thin-film systems, we observe that the bandpass filter selection method yields results comparable to those obtained for the vdW materials (see Figure~S6). In the case of GaP on \ch{SiO2}, we find a minimum distinguishability of $D_\mathrm{min}^\mathrm{GaP} = 0.076$, which is only slightly lower than the value obtained for 3R-\ch{MoS2}, despite the substantially lower refractive index contrast of GaP ($n \approx 3.1$) on \ch{SiO2} ($n \approx 1.44$) when compared to 3R-\ch{MoS2} ($n \approx 4$) on PDMS ($n \approx 1.4$).

Similarly, for Si on \ch{SiO2}, we obtain a minimum distinguishability of $D_\mathrm{min} = 0.089$, which is the highest value among all studied systems. 
The refractive index of Si, $n \approx 3.5$, lies between those of GaP and 3R-\ch{MoS2}.
This suggests that there is a broad range of dielectric functions for which the thickness characterization technique performs close to optimally over our chosen target interval up to \SI{1}{\um}.

Despite the significantly lower refractive index value of \ch{SiO2} ($n \approx 1.44$) in comparison to hBN ($n \approx 2.2$), we obtain only a slightly diminished $D_\mathrm{min}$ value of $D_\mathrm{min}^\mathrm{SiO2} = 0.040$ for the \ch{SiO2} on Si thin-film system. This is due to the refractive index contrast not being limited by the substrate, with $n \approx 3.5$, but by the cladding medium, with $n = 1$, which results in a comparable relative contrast of approximately $50\%$ and similar oscillation amplitudes in reflectance with respect to film thickness.

Remarkably, constraining one of the bandpass filters to $\lambda_\mathrm{c} = \SI{1650}{\nm}$ did not improve the optimization result, as shown in Figure~S8. This indicates that, unlike for higher-index materials, the thickness characterization accuracy for \ch{SiO2} is primarily limited by the reflectance oscillation amplitude rather than the oscillation period. Thus, the target thickness interval could likely be expanded for the set of bandpass filters considered here, without a substantial decrease in $D_\mathrm{min}$.

Altogether, these numerical analyses highlight the broad applicability of the proposed multispectral imaging approach. The set of bandpass filters can be tailored to the refractive index contrast and transparency window of a given material-substrate system. This allows robust, high-precision, spatially resolved thickness mapping to be extended to a wide variety of thin-film systems and positions our technique not merely as a specialized solution for pre-screening vdW flakes on polymer substrates, but as a highly adaptable, non-destructive, and cost-effective tool for general optical thin-film metrology.

\section{Conclusion}
We have presented a rigorous framework for the thickness characterization of optical thin films using sparse spectral sampling that combines theoretical modeling with experimental refinement and incorporates statistical correlations between measurement channels to enhance robustness against observed systematic deviations. The method’s key advantages are its simplicity, scalability, and compatibility with polymer substrates. It therefore addresses a critical limitation of conventional techniques, which typically require flat, rigid surfaces. The hierarchical confidence metrics ($C_\mathrm{G}$, $C_\mathrm{L}$, and $\Delta t_R$) provide built-in quality control, enabling the identification of potentially misclassified regions in thickness maps. 

We have demonstrated the technique's practical applicability for the rhombohedral (3R) polytype of the transition metal dichalcogenide molybdenum disulfide on polydimethylsiloxane (PDMS) substrates. By strategically selecting five discrete near-infrared wavelengths, we achieved unambiguous thickness determination up to \SI{691}{\nm} with a mean 95\% confidence-interval width of \SI{8.3}{\nm}.

Furthermore, we have shown that the bandpass filter selection strategy can be extended to materials with a broad range of dielectric functions. This suggests wide applicability across commonly used optical materials and also highlights the potential of the method as a general tool for ultrafast thin film characterization in fabrication workflows.

\section{Methods} \label{sec:meth}

\subsection{Sample Preparation}
3R-\ch{MoS2} flakes were mechanically exfoliated from a commercially acquired bulk crystal (HQ Graphene) using the standard Scotch tape method. The exfoliated flakes were transferred onto polydimethylsiloxane (PDMS) substrates (Gel-Pak, PF-30-X4) by pressing the tape onto the PDMS surface and peeling it off. Each PDMS sample was mounted onto a glass microscope slide while keeping the protective backside cover attached to maintain substrate flatness during imaging.

\subsection{Multispectral Imaging} \label{sec:meth:msi}
Reflectance measurements were performed using a custom-built near-infrared (NIR) microscope setup. A broadband halogen lamp (Thorlabs SLS302) served as the illumination source, providing continuous spectral output from \SI{360}{\nm} to \SI{2500}{\nm}. The light was reflected off a 50:50 beamsplitter before being focused onto the sample using a reflective Schwarzschild objective (Thorlabs LMM40X-P01, NA = 0.5, 40x magnification). The reflected light was collected through the same objective, passed through the beamsplitter, and directed toward the detection path.

Spectral filtering was achieved using a set of five commercially available bandpass filters (Thorlabs FBH920-10, FBH1100-10, FBH1350-12, FBH1480-12, and FBH1650-12). The filters were mounted on a motorized filter wheel (Thorlabs FW102C) for automated sequential imaging. Images were captured using an InGaAs camera (Xenics Xeva 320 Series, XEN-000101) with a 320 $\times$ 240 pixel sensor and a spectral sensitivity up to \SI{1700}{\nm}.

To obtain absolute reflectance values, reference images were acquired using a protected silver mirror (Thorlabs PF10-03-P01). The sample reflectance $R$ was calculated as:
\begin{align}
	R = \frac{I_\mathrm{sample} - I_\mathrm{dark}}{I_\mathrm{mirror} - I_\mathrm{dark}}
\end{align}
where $I_\mathrm{sample}$, $I_\mathrm{mirror}$, and $I_\mathrm{dark}$ are the measured intensities of the sample, mirror, and dark signal, respectively. The sample and reference images for each bandpass filter channel were acquired with identical illumination conditions and exposure times.

\subsection{Vertical Scanning Interferometry (VSI)}
Following multispectral imaging, each sample was transferred onto a flat fused silica substrate in a dry stamping procedure.
The VSI measurements were performed using a white light optical surface profiler (Bruker Contour GT) in VXI mode with a 50× Mirau objective. The acquired height maps were post-processed using the open-source data analysis software Gwyddion \cite{necasGwyddionOpensourceSoftware2012} to correct for sample tilt and background offset using the three-point plane-fitting and subtraction method.

\subsection{Data Processing and Analysis} \label{sec:data_proc}
All image processing and data analysis were performed with custom Python scripts. An affine transformation in pixel/data coordinates was used to correlate each multispectral reflectance image acquired before transfer with the corresponding VSI measurement acquired after transfer. The transformation is given by:
\begin{align}
    \begin{pmatrix}
        x' \\
        y'
    \end{pmatrix} = \alpha
    R(\theta) \begin{pmatrix}
        x - x_0 \\
        y - y_0
    \end{pmatrix}.
\end{align}
Here, $x$, $y$, $x'$, and $y'$ are the coordinates in the respective measurements, $R(\theta)$ is the rotation matrix for angle $\theta$, $x_0$ and $y_0$ are offsets in their respective coordinate directions, and $\alpha$ is the scaling between the two coordinate systems.
For each characterized sample, we manually selected distinct pairs of corresponding points in the reflectance and VSI maps. The optimal transformation parameters were then obtained by least-squares minimization of the total distance mismatch across all point pairs.

\subsection{Transfer Matrix Method} \label{sec:meth:tmm}
We modeled the reflectance response of 3R-\ch{MoS2} (and all other materials) using a transfer matrix approach. The transfer matrix of a single layer with refractive index $n$ and thickness $t$ is given by:
\begin{align}
    M = \begin{pmatrix}
        \cos(k_z t) & \sin(k_z t) / k_z \alpha \\
        - k_z \alpha \sin(k_z t) & \cos(k_z t)
    \end{pmatrix}
\end{align}
where $k_z$ is the component of the wavevector along the layer-normal direction $z$, and $\alpha = 1$ for TE-polarized light and $\alpha = 1 / n^2$ for TM-polarized light.

For TM-polarized light, the uniaxial anisotropy of 3R-\ch{MoS2} must be taken into account. In this case, $n$ is calculated from the ordinary, or in-plane, refractive index $n_\mathrm{o}$, the extraordinary, or out-of-plane, refractive index $n_\mathrm{eo}$, and the propagation angle inside the material, $\theta_\mathrm{MoS2}$, relative to the ordinary axis of the crystal \cite{wagonerSecondharmonicGenerationMolybdenum1998}:
\begin{align}
    n = \left( \frac{\cos^2(\theta_\mathrm{MoS2})}{n_\mathrm{o}^2} + \frac{\sin^2(\theta_\mathrm{MoS2})}{n_\mathrm{eo}^2} \right)^{-\frac{1}{2}}
\end{align}
for which $\theta_\mathrm{MoS2}$ is obtained from a generalized form of Snell's law based on the angle of incidence $\theta_\mathrm{AOI}$ \cite{wagonerSecondharmonicGenerationMolybdenum1998}:
\begin{align}
   \theta_\mathrm{MoS2} = \arctan \left( \frac{\theta_\mathrm{AOI} / n_\mathrm{o}^2}{1 - \sin^2(\theta_\mathrm{AOI})/ n_\mathrm{eo}^2} \right)^{\frac{1}{2}}
\end{align}
where we assume $n = 1$ for the refractive index of the medium from which the light is incident.

Based on $M$, the refractive index of the substrate $n_\mathrm{sub}$ (PDMS), and the $z$-components of the wavevector of the light entering and exiting the layer, $k_z^\mathrm{inc}$ and $k_z^\mathrm{sub}$, the reflection coefficients for TE and TM polarization are given by
\begin{align}
    r_\mathrm{TE} &= \frac{(k_z^\mathrm{inc} M_{22} - k_z^\mathrm{sub} M_{11}) - i (M_{21} + k_z^\mathrm{inc} k_z^\mathrm{sub} M_{12})}{(k_z^\mathrm{inc} M_{22} + k_z^\mathrm{sub} M_{11}) + i (M_{21} - k_z^\mathrm{inc} k_z^\mathrm{sub} M_{12})} \\
    r_\mathrm{TM} &= \frac{(n_\mathrm{sub}^2 k_z^\mathrm{inc}  M_{22} - k_z^\mathrm{sub} M_{11}) - i (n_\mathrm{sub}^2 M_{21} + k_z^\mathrm{inc} k_z^\mathrm{sub} M_{12})}{(n_\mathrm{sub}^2 k_z^\mathrm{inc}  M_{22} + k_z^\mathrm{sub} M_{11}) + i (n_\mathrm{sub}^2 M_{21} - k_z^\mathrm{inc} k_z^\mathrm{sub} M_{12})}
\end{align}
respectively. The corresponding reflectances are then obtained by calculating $R_\mathrm{TE/TM} = |r_\mathrm{TE/TM}|^2$. The optical transfer matrix calculations were implemented in Python.

\subsection{Schwarzschild Objective Reflectance Modeling} \label{sec:ss_obj_modeling}
To model the observed reflectance response for each bandpass filter, we calculated the reflectance as a function of thickness $t$ and angle of incidence $\theta$ for TE- and TM-polarized light for the corresponding center wavelength $\lambda_\mathrm{c}$ with the transfer matrix method described above. Both polarizations were averaged and, assuming uniform back-focal plane illumination of the Schwarzschild objective and negligible spatial coherence of our light source, weighted with $\sin(\theta)$ for each incidence angle $\theta$:
\begin{align}
	R_{\lambda_\mathrm{c}}(t) = \frac{1}{2N}\int_{\theta_\mathrm{min}}^{\theta_\mathrm{max}} \left( R_\mathrm{TE}(\lambda_\mathrm{c}, t, \theta) + R_\mathrm{TM}(\lambda_\mathrm{c}, t, \theta) \right) \sin(\theta) \mathrm{d} \theta.
\end{align}
$\theta_\mathrm{max} = \SI{30}{\degree}$ was derived from the numerical aperture of the objective and $\theta_\mathrm{min} = \SI{12}{\degree}$ from the diameter of the obstructing mirror. Both quantities were taken from manufacturer specifications. Note that we approximated the response of each bandpass filter by its center wavelength. Furthermore, the normalization factor $N$ was calculated as:
\begin{align}
	N = \int_{\theta_\mathrm{min}}^{\theta_\mathrm{max}} \sin(\theta) \mathrm{d} \theta.
\end{align}

\section*{Acknowledgments}
The authors acknowledge the support from the Deutsche Forschungsgemeinschaft (DFG, German Research Foundation), Project ID: 437527638—IRTG 2675 (Meta-Active), and Collaborative Research Center (CRC/SFB) 1375 NOA Project B3.

\section*{Disclosures}
The authors declare no conflicts of interest.

\section*{Data Availability}
Data underlying the results presented in this paper are not publicly available at this time but may be obtained from the authors upon reasonable request.

%%%%%%%%%%%%%%%%%%%%%%% References %%%%%%%%%%%%%%%%%%%%%%%%%
\printbibliography

\newpage
\appendix
\noindent{\LARGE{\textsf{\textbf{Supplementary}}}}

% Sections: S1, S2, S3, ...
\renewcommand{\thesection}{S\arabic{section}}
\renewcommand{\thesubsection}{S\arabic{section}.\arabic{subsection}}
% Figures: S1, S2, S3, ...
\renewcommand{\thefigure}{S\arabic{figure}}
% Tables: S1, S2, S3, ...
\renewcommand{\thetable}{S\arabic{table}}
% Equations: S1, S2, S3, ...
\renewcommand{\theequation}{S\arabic{equation}}
% Reset counters so numbering starts from S1
\setcounter{section}{0}
\setcounter{figure}{0}
\setcounter{table}{0}
\setcounter{equation}{0}

\section{VSI Measurements of 3R-\ch{MoS2} on PDMS}
\begin{figure}[htbp]
	\centering
	\includegraphics{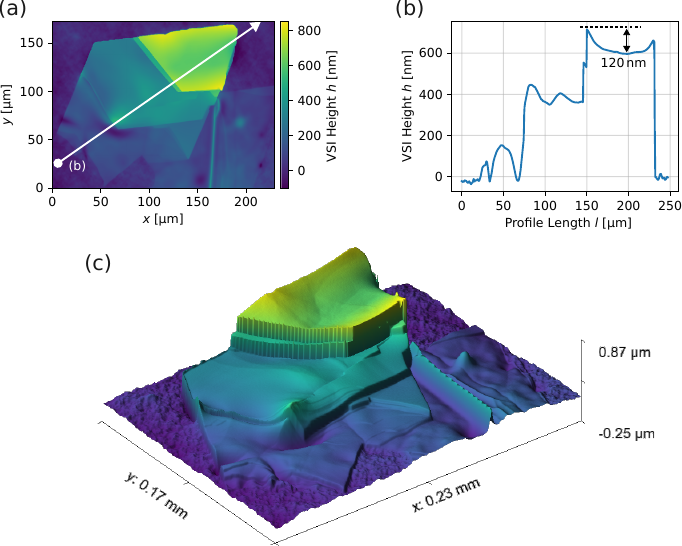}
	\caption{\textbf{Vertical scanning interferometry (VSI) measurement of a 3R-\ch{MoS2} flake on PDMS.}
	\textbf{(a)} Obtained height map.
	\textbf{(b)} Height profile along the long axis of the flake as indicated in (a), showing height fluctuations and measurement distortions.
	\textbf{(c)} Shaded 3D representation of the VSI measurement showing an apparent indentation of the flake on the thickest section.}
	\label{fig:vsi_on_pdms}
\end{figure}

As stated in the main text, conventional thickness characterization techniques, such as vertical scanning interferometry (VSI), produce inaccurate results for measurement on PDMS, as any substrate deformations directly translate to film deformations, that cannot be corrected for without precise knowledge of the substrate geometry. This is demonstrated in Figure~\ref{fig:vsi_on_pdms} where the results of a VSI measurement of a 3R-\ch{MoS2} flake on a PDMS substrate are displayed.

The height mapping in Figure~\ref{fig:vsi_on_pdms}(a) represents the best estimate for material thickness, which we obtained by leveling the data by subtracting a plane of height values fitted to the freely exposed PDMS substrate.
Taking the profile along the long axis of the flake, as annotated in Figure~\ref{fig:vsi_on_pdms}(a) and displayed in Figure~\ref{fig:vsi_on_pdms}(b), we see that, while the approximate shape of the flake can be discerned, precise and consistent evaluation of the material's thickness is not possible, due the distortion introduced by bonding to the PDMS substrate. Note that this is especially pronounced between profile lengths of $l = \SI{150}{\um}$ and $l = \SI{200}{\um}$ where a height difference of \SI{120}{\nm} is observed, despite the flake material itself lacking any steep changes in height, characteristic for thickness boundaries in layered materials. Note that this apparent indentation is also visible in the shaded 3D representation of the flake displayed Figure \ref{fig:vsi_on_pdms}(c).

\section{Thickness Dependence of the Quality Metrics}
The reflectance based thickness method presented in the main text provides intrinsic limits for the three derived quality metrics global confidence $C_\mathrm{G}$, local confidence $C_\mathrm{L}$, and precision $\Delta t_R$.

\begin{figure}[htbp]
    \centering
    \includegraphics[scale=0.6]{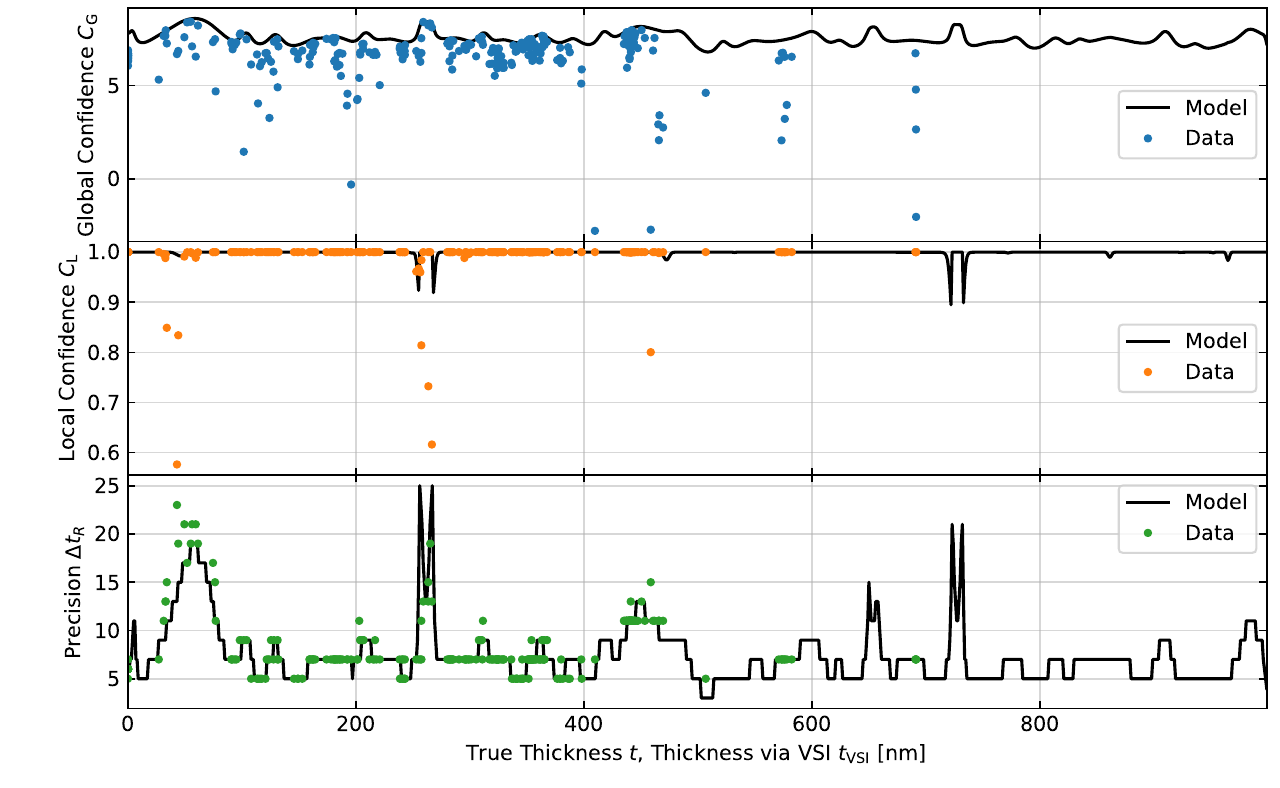}
    \caption{Model intrinsic upper limits for global confidence $C_\mathrm{G}$, local confidence $C_\mathrm{L}$, and precision $\Delta t_R$ and observed values as a function of thickness.}
    \label{fig:model_intrinsic_quality_metrics}
\end{figure}

\begin{figure}[htbp]
    \centering
    \includegraphics[scale=0.6]{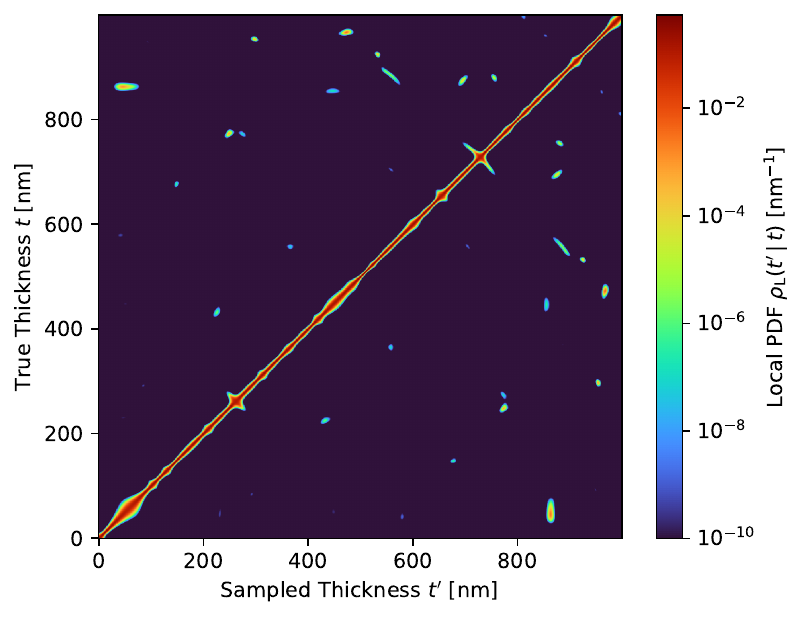}
    \caption{Model intrinsic local probability density function $\rho_\mathrm{L}(t' | t)$ of sampled thickness $t'$ for any given true thickness $t$. $\rho_\mathrm{L}$ is fundamentally symmetric under permutation of $t$ and $t'$.}
    \label{fig:local_PDF_mat}
\end{figure}

To study the model intrinsic, thickness dependent behavior of $C_\mathrm{G}$, $C_\mathrm{L}$, and $\Delta t_R$, we calculate all possible residual combinations within in the studied interval of thickness values up to \SI{1}{\um}
\begin{align}
    \delta \vec{R}(t', t) = \vec{R}_\mathrm{model}(t') - \vec{R}_\mathrm{model}(t),
\end{align}
where $\vec{R}_\mathrm{model}$ is the refined reflectance model shown in Figure 3(a) of the main text, and for the sake of simplicity, we treat $t$ and $t'$ as the true and sampled thickness values, respectively.
Performing the thickness retrieval procedure as described in the main text, we obtain the quality metrics as a function of thickness, which are displayed as black lines in Figure \ref{fig:model_intrinsic_quality_metrics}. 
Additionally, we obtain the local probability density function (PDF) matrix in Figure \ref{fig:local_PDF_mat}. Note that, in Figure \ref{fig:model_intrinsic_quality_metrics} we have also plotted the experimentally observed quality metrics obtained from the thickness evaluation of the model refinement data set.

For the global confidence, the model intrinsic values undulate between $C_\mathrm{G} = 6.8$ and $C_\mathrm{G} = 8.6$, while the observed values are above $C_\mathrm{G} = 5$, in the vast majority of cases.
The model intrinsic local confidence is close to unity for almost the entire studied thickness range. The only exceptions are the two narrow double peak structures around $t = \SI{262}{\nm}$ and $t = \SI{728}{\nm}$, which are caused by the crests perpendicular to  the main diagonal of the local PDF matrix $\rho_\mathrm{L}(t'| t)$ at those thickness values (see Figure \ref{fig:local_PDF_mat}).
For $\Delta t_R$ we observe a strong variation with respect to thickness throughout the target interval, with $\min(\Delta t_R) = \SI{3}{\nm}$ and $\max(\Delta t_R) = \SI{25}{\nm}$. Note that the step like structure is due to the $\SI{1}{\nm}$ resolution that was chosen for thickness sampling.
Furthermore, drops in precision are observed around the previously mentioned drops in $C_\mathrm{L}$, as $\Delta t_R$ similarly spikes in a double peak like structure. An additional feature of note in $\Delta t_R$ is the broad peak around $t = \SI{58}{\nm}$, which is caused by a broad section of the main diagonal of $\rho_\mathrm{L}(t'| t)$.

\section{Thickness Characterization Results of a Separate Dataset}
\begin{figure}[htbp]
    \centering
    \includegraphics{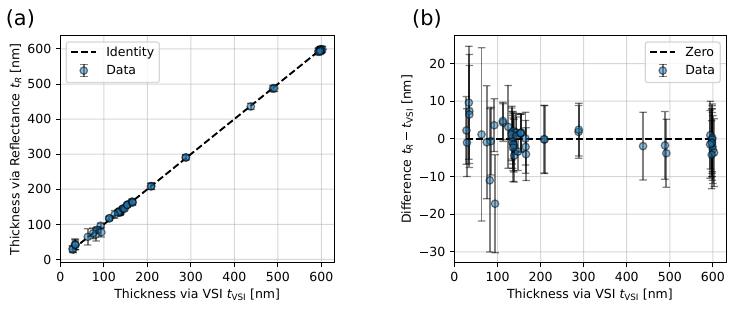}
    \caption{ \textbf{Thickness determination of 3R-\ch{MoS2} on PDMS samples not used for model refinement.}
    	(a) Thickness as characterization through the BPF based reflectance method plotted against the thickness as determined via VSI after transfer onto a glass substrate. The errorbars denote the precision $\Delta t_R$. (b) Difference between both measurements. The errorbars also denote $\Delta t_R$.}
    \label{fig:non_refinement_dset}
\end{figure}

We also applied the the thickness characterization method to a dataset of VSI and reflectance measurements which were not used in the model refinement process. Masking out any data-points which violate the heuristically obtained limits for $C_\mathrm{G}$, $C_\mathrm{L}$, and $\Delta t_R$, we obtain the results displayed in Figure~\ref{fig:non_refinement_dset}(a) and (b) which show that, in the vast majority of cases, $t_R$ is in agreement with $t_\mathrm{VSI}$ within the determined precision $\Delta t_R$

\section{Other Materials and Thin Film Systems}
\begin{figure}[htbp]
    \centering
    \includegraphics{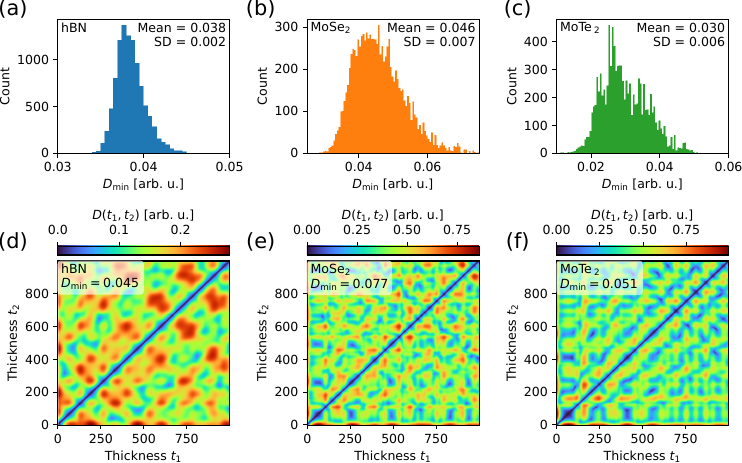}
    \caption{\textbf{Bandpass filter optimization results for other vdW materials on PDMS.}
    	(a)-(c) Histogram of $D_\mathrm{min}$ values obtained via the optimization procedure for hBN, \ch{MoSe2}, and \ch{MoTe2}, respectively. Mean and standard deviation (SD) of each histogram are given as annotated text. (d)-(f) Distinguishability matrix $D(t_1, t_2)$ for the best set of bandpass filters for each material.}
    \label{fig:BPF_select_other_vdW}
\end{figure}

As described in the main text, we also performed the bandpass filter selection scheme for the van der Waals (vdW) materials hexagonal boron nitride (hBN), molybdenum diselenide (\ch{MoSe2}), and molybdenum ditelluride (\ch{MoTe2}).
The resulting optimization histograms, displaying the counts of minimum separation values $D_\mathrm{min}$ obtained during $N = 10000$ attempts, are shown in Figure \ref{fig:BPF_select_other_vdW}(a)-(c), respectively, for each material.
In Figure \ref{fig:BPF_select_other_vdW}(d)-(f), the distinguishability matrix for the optimal (highest $D_\mathrm{min}$) set of bandpass filters is displayed for each respective material.

\begin{figure}[htbp]
    \centering
    \includegraphics{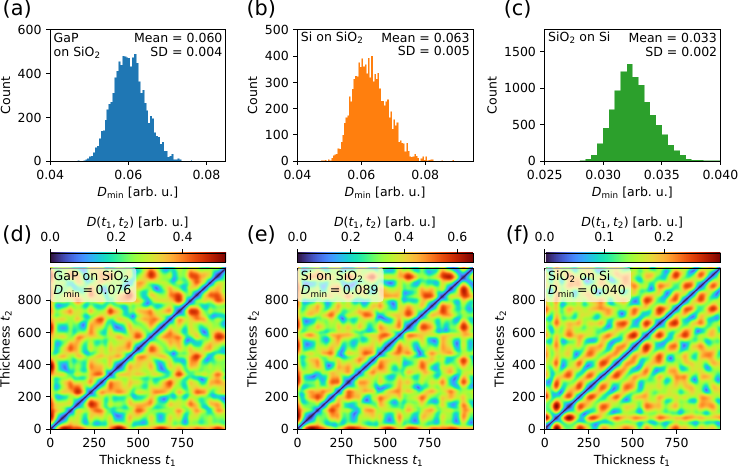}
    \caption{\textbf{Bandpass filter optimization results for conventional thin-film systems.}
    	(a)-(c) Histogram of $D_\mathrm{min}$ values obtained via the optimization procedure for \ch{GaP} on \ch{SiO2}, \ch{Si} on \ch{SiO2}, and \ch{SiO2} on \ch{Si}, respectively. Mean and standard deviation (SD) of each histogram are given as annotated text. (d)-(f) Distinguishability matrix $D(t_1, t_2)$ for the best set of bandpass filters for each material.}
    \label{fig:BPF_select_conv_materials}
\end{figure}

Similarly, we applied the optimization procedure to conventional thin-film systems, specifically gallium phosphide (\ch{GaP}) on silicon dioxide (\ch{SiO2}), silicon (\ch{Si}) on \ch{SiO2}, and \ch{SiO2} on \ch{Si}. The resulting histograms of $D_\mathrm{min}$ values and the distinguishability matrices for the best bandpass filter sets are shown in Figure~\ref{fig:BPF_select_conv_materials}.

\begin{figure}[htbp]
    \centering
    \includegraphics[scale=0.6]{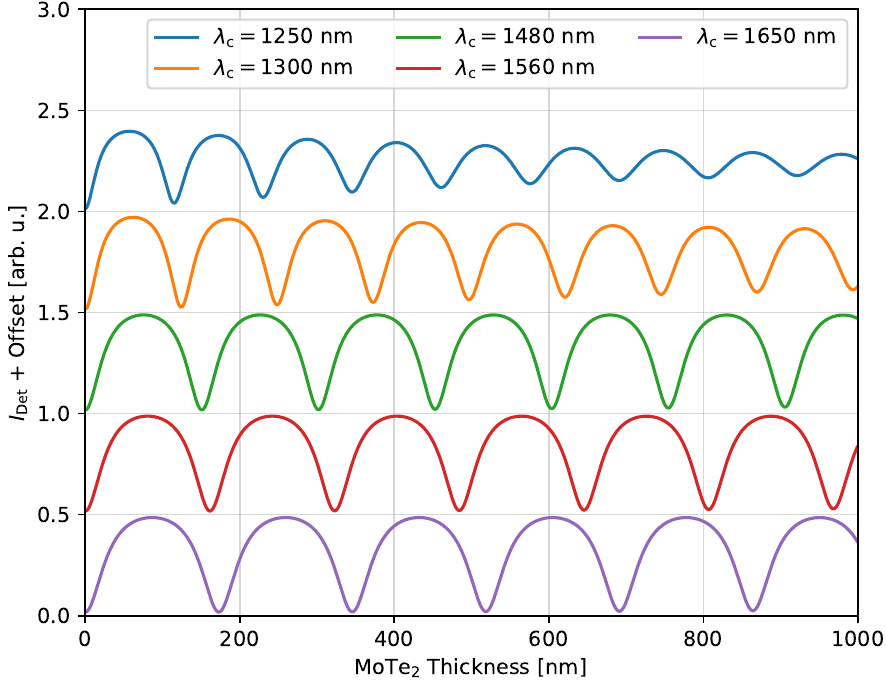}
    \caption{Reflectance as a function of thickness for the optimal bandpass filter set
    $\Lambda = \{ \SI{1250}{\nm}, \SI{1300}{\nm}, \SI{1480}{\nm}, \SI{1500}{\nm}, \SI{1500}{\nm}, \SI{1650}{\nm} \}$ for thickness determination of \ch{MoTe2} on PDMS.
    For the center wavelengths $\lambda_\mathrm{c}$ below the absorption cut-off at $\lambda = \SI{1350}{\nm}$, dampening of the oscillation amplitude is observed with increasing thickness.}
    \label{fig:mote2_opt_refl}
\end{figure}

Figure~\ref{fig:mote2_opt_refl} shows the reflectance of \ch{MoTe2} as a function of thickness for the optimal bandpass filter set. Due to the material's absorption cut-off at $\lambda \approx \SI{1350}{\nm}$, dampening of the oscillation amplitude is observed for center wavelengths $\lambda_\mathrm{c}$ below this value.

\begin{figure}[htbp]
    \centering
    \includegraphics[scale=0.6]{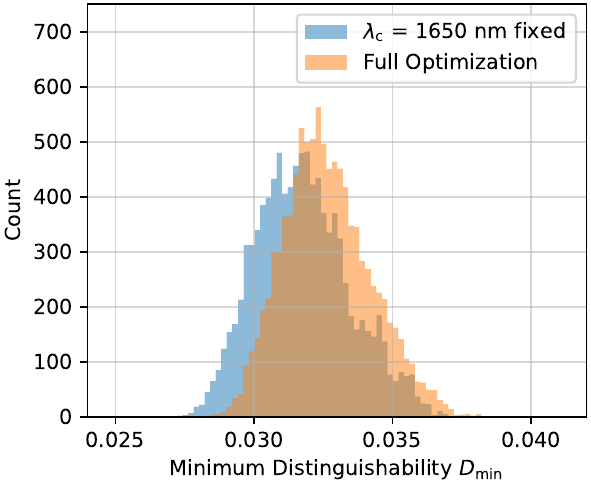}
    \caption{Bandpass filter optimization histograms for an \ch{SiO2} on \ch{Si} layer system constraining one of bandpass filters to the longest center wavelength $\lambda_\mathrm{c} = \SI{1650}{\nm}$ (blue) and a fully unconstrained optimization (orange). Contrary to the results of the other tested materials with higher refractive indices, constraining the optimization yields lower $D_\mathrm{min}$ for the chosen target interval of thicknesses up to \SI{1}{\um}.}
    \label{fig:sio2_on_si_opt_hists}
\end{figure}

Finally, Figure~\ref{fig:sio2_on_si_opt_hists} presents the bandpass filter optimization histograms for an \ch{SiO2} on \ch{Si} layer system. The figure compares a fully unconstrained optimization (orange) with an optimization where one of the bandpass filters is constrained to the longest center wavelength $\lambda_\mathrm{c} = \SI{1650}{\nm}$ (blue). For this material system, and the chosen target interval of thicknesses up to \SI{1}{\um}, constraining the optimization yields lower $D_\mathrm{min}$ values.

\end{document}